\documentclass[a4paper,pre,reqno,superscriptaddress,twocolumn, floatfix]{revtex4}
\usepackage{graphicx,color}
\usepackage{dcolumn}
\usepackage{epsfig}  
\usepackage[centertags]{amsmath}
\usepackage{amsfonts}
\usepackage{euscript}
\usepackage{amssymb}
\usepackage{amsthm}
\usepackage{newlfont}
\usepackage{lipsum}
\usepackage{mathtools}
\usepackage{mathrsfs}
\usepackage{subfigure}
\usepackage{todonotes}
\usepackage{kbordermatrix}
\usepackage[normalem]{ulem}
\usepackage[utf8]{inputenc}
\usepackage{comment}
\usepackage{float}
\usepackage{lipsum}  
\usepackage[colorlinks=true,linkcolor=blue,urlcolor=blue,citecolor=blue,bookmarks=true]{hyperref}
\usepackage{nameref}
\usepackage{enumitem}
\usepackage{natbib}
\usepackage{soul}
\usepackage{xcolor}

\begin{document}

\title{Drivers of cooperation in social dilemmas on higher-order networks}

\author{Onkar Sadekar}
\affiliation{Department of Network and Data Science, Central European University Vienna, Vienna 1100, Austria}

\author{Andrea Civilini}
\affiliation{School of Mathematical Sciences, Queen Mary University of London, London E1 4NS, United Kingdom}
\affiliation{Dipartimento di Fisica ed Astronomia, Universit\`a di Catania and INFN, Catania I-95123, Italy}

\author{Vito Latora}
\affiliation{School of Mathematical Sciences, Queen Mary University of London, London E1 4NS, United Kingdom}
\affiliation{Dipartimento di Fisica ed Astronomia, Universit\`a di Catania and INFN, Catania I-95123, Italy}
\affiliation{Complexity Science Hub Vienna, A-1080 Vienna, Austria}

\author{Federico Battiston}
\affiliation{Department of Network and Data Science, Central European University Vienna, Vienna 1100, Austria}

\begin{abstract}
Understanding cooperation in social dilemmas requires models that capture the complexity of real-world interactions.
While network frameworks have provided valuable insights to model the evolution of cooperation, they are unable to encode group interactions properly. 
Here, we introduce a general higher-order network framework for multi-player games on structured populations. 
Our model considers multi-dimensional strategies, based on the observation that social behaviours are affected by the size of the group interaction. 
We investigate dynamical and structural coupling between different orders of interactions, revealing the crucial role of nested multilevel interactions, and showing how such features can enhance cooperation beyond the limit of traditional models with uni-dimensional strategies.
Our work identifies the key drivers promoting cooperative behaviour commonly observed in real-world group social dilemmas.
\end{abstract}

\maketitle

\section{Introduction}

%general human behaviour
One remarkable characteristic observed across many species is their ability to form intimate bonds with non-related individuals and sustain large-scale cooperation in complex social groups \cite{smith_logic_1973, nowak_supercooperators_2012, christakis_friendship_2014}. In contrast, the Darwinian principle of natural selection dictates that self-interested individuals should avoid engaging in altruistic actions, as the costs of such cooperative behaviours do not necessarily confer a direct reproductive advantage \cite{hardin_tragedy_1968, maynard_smith_evolution_1982, dawkins_selfish_2006}. The same principle extends also to human social systems. For example, in a collective action problem, when given a choice between maximizing personal gains or improving societal benefits at a personal cost, a purely rational approach would dictate that individuals defect against one another \cite{taylor_evolutionary_1978, hofbauer_evolutionary_1998}.

%game theory and evolutionary game theory
Understanding the emergence of cooperative behaviours in such seemingly unfavourable situations has been one of the long-standing goals of natural and social sciences \cite{nowak_evolutionary_1992, szabo_evolutionary_1998}. A pivotal turning point in approaching this problem was the development of game theory, a mathematical framework for analyzing strategic interactions \cite{von1944theory, nash_equilibrium_1950, nowak_evolving_2012, broom_generalized_2019}. Over the last century, game theory has undergone significant advancements and has been extensively applied to the study of so-called social dilemmas -- scenarios that involve a fundamental tension between individual choice and the collective good \cite{axelrod_evolution_1981, axelrod_further_1988, sigmund_evolutionary_1999, nowak_evolutionary_2006, szabo_evolutionary_2007, perc_statistical_2017}.

Usually, social dilemmas comprise of two interacting individuals, each of whom earns a payoff based on their strategies, typically denoted as cooperation (C) and defection (D). Combining game theory with the Darwinian principle of the \emph{survival of the fittest}, evolutionary game theory extends this paradigm to consider repeated interactions in a population. The fraction of individuals adopting each strategy changes over time based on how fit the strategy is in the population, where the fitness is determined by the average payoff earned by players adopting that particular strategy \cite{axelrod_evolution_1981, nowak_evolutionary_1992, hofbauer_evolutionary_1998, traulsen_future_2023}. Over the years, extensive numerical and theoretical work for various games and social dilemma situations have identified several different mechanisms contributing to the success of cooperative behaviour. These mechanisms can be broadly classified into reciprocity based (direct \cite{trivers1971evolution, axelrod_evolution_1981}, indirect \cite{nowak1998evolution, hilbe2018indirect}, or network \cite{lieberman_evolutionary_2005, szabo_evolutionary_2007}) or selection based (kin \cite{hamilton1964genetical, taylor1992altruism} or group \cite{wilson1975theory, traulsen2006evolution}) as pointed out in \cite{nowak_five_2006}.

%games on networks
Repeated interactions with the same set of individuals (or neighbours) is at the heart of network reciprocity which has gathered substantial attention in recent years as a computational and mathematical framework able to explain cooperation in socio-structural interaction patterns \cite{hofbauer_evolutionary_1998, santos_scale-free_2005, lieberman_evolutionary_2005, nowak_five_2006, santos_evolutionary_2006, szabo_evolutionary_2007, gomez-gardenes_dynamical_2007}. In particular, the effect of different interaction structures has been investigated, from lattices \cite{nowak_evolutionary_1992, szabo_evolutionary_2007, kumar_evolution_2020}, to scale-free networks \cite{santos_scale-free_2005, santos_social_2008}, and multi-layer networks \cite{wang_evolutionary_2015, battiston_determinants_2017,su_evolution_2022} enhancing our understanding of how pro-sociality evolves in large-scale structured populations with realistic features. A crucial limitation of networks is that they encode ties with links, and are hence unable to properly represent non-pairwise interactions. By contrast, group sociality is a ubiquitous feature of human interaction, highlighting the need for frameworks that can accommodate multi-player interactions within structured populations \cite{broom_multi-player_1997, tarnita_set_game_2009, gokhale_evolutionary_2010, broom_generalized_2019}.

%games on hypergraphs
Hypergraphs (or higher-order networks) are mathematical objects that encode group interactions through hyperedges, describing interactions among an arbitrary number of individuals \cite{battiston_networks_2020, battiston_physics_2021}. A variety of dynamical processes such as spreading of behaviours \cite{iacopini_simplicial_2019}, synchronization of coupled oscillators \cite{skardal_higher_2020, gambuzza_stability_2021,zhang_higher-order_2023}, and very recently evolutionary games \cite{civilini_evolutionary_2021, alvarez-rodriguez_evolutionary_2021, guo_evolutionary_2021, xu2024reinforcement}, have been shown to exhibit fundamentally different behaviour when considered on hypergraphs. In particular, a higher-order prisoner's dilemma game was shown to exhibit an explosive transition to cooperation at a critical fraction of higher-order interactions \cite{civilini_explosive_2024}. Similar works on higher-order networks for other types of games such as public goods game \cite{wang_evolutionary_2024} and sender-receiver game \cite{kumar_evolution_2021} have shown analogous non-trivial behaviour with respect to the topological properties of the hypergraphs.

%limitations and literature gap
Despite these advances, the proposed models suffer from a variety of limitations. First, most of them consider that the agents choose the same strategy across groups of different sizes. Empirical research has shown that this is usually not the case \cite{krause_social_2007, funato_model_2011, cantor_primer_2020, ogino_group-level_2023, quan_cooperation_2024}. Larger groups can exert peer pressure which does not necessarily grow linearly with group size \cite{capraro_group_2015, pena_group_2018, pereda_group_2019}. Additionally,  (pairwise) edges and hyperedges are usually treated as either independent from each other (random hypergraphs) or completely overlapping (simplicial complexes) \cite{battiston_networks_2020, zhang_higher-order_2023, malizia2025hyperedge}. Even though these constraints ease the mathematical difficulty for analytical calculations, multiple social interaction datasets consisting of higher-order relations have shown that real-world network structures might lay in between such two extreme cases \cite{zhang_higher-order_2023, gallo_higher-order_2024, iacopini_temporal_2024, ma_optimal_2024}.

In this work, we introduce a new multi-dimensional strategic model for games on hypergraphs inspired by evolutionary processes on multiplex networks \cite{battiston_determinants_2017}. We introduce our model consisting of order-dependent strategies. We then investigate the dynamical coupling between strategic behaviours at different orders and propose a simple tunable model of random hypergraphs to reveal the effect of structural overlap on prosocial behaviour. Finally, we investigate how the strength of higher-order social dilemmas affects the emergence of cooperation in networked populations.

\begin{figure*}[!ht]
    \centering
    \includegraphics[width=\textwidth]{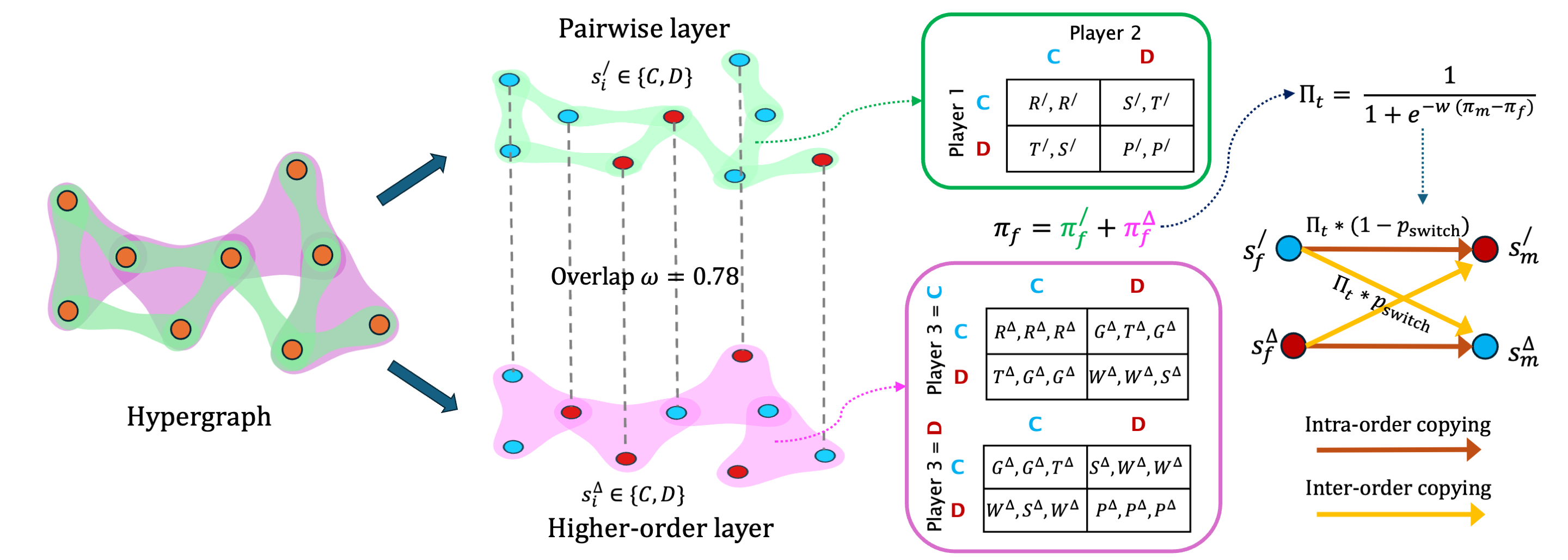}
    \caption{\textbf{Evolutionary dynamics of higher-order games.} Hypergraph consisting of pairwise (green) and higher-order (pink) interactions connects individuals. A tunable fraction $\omega$ describes the structural overlap between the different orders. Each individual has order-specific strategies ($s_i^/$ and $s_i^{\Delta}$) to either cooperate (C) or defect (D) with others and plays the corresponding 2 or 3-player games with its neighbours. After accumulating the payoff across both orders, the player imitates the strategy in either order of a random neighbour based on the dynamical coupling $p_{\text{switch}}$.}
    \label{fig:illustration}
\end{figure*}

\section{Model}\label{sec:model}

A population of $N$ players is represented as a hypergraph $\mathcal{H}(\mathcal{V}, \mathcal{E})$, where the players are the nodes of the hypergraph $\mathcal{V}$, such that $|\mathcal{V}|=N$ and the set of hyperedges $\mathcal{E}$ represents the set of games played in the population.
Hyperedges are a generalization of the network's edges to group interactions, a hyperedge representing a group of an arbitrary number of interacting nodes.
In particular, we focus on the case where the hypergraph $\mathcal{H}$ consists of only $2-$hyperedges and $3-$hyperedges, respectively denoted by $\mathcal{E}^/$  and $\mathcal{E}^{\Delta}$. A $2$-player game is associated to each $2-$hyperedge, while a $3$-player game to each $3-$hyperedge. Henceforth, we will use superscripts $/$ and $\Delta$ to indicate pairwise and higher-order quantities respectively, while reserving the subscripts for node labels.
Each player participates in all the hyperedges to which it belongs. Since every player is part of multiple hyperedges, they engage in a separate game for each one. Thus, the total number of games a player $i$ takes part in equals their hyperdegree $k_i$. We define $k_i^/$ and $k_i^{\Delta}$  as the number of 2-player and 3-player games player $i$ is involved in, respectively, satisfying $k_i^/ + k_i^{\Delta} = k_i$.

Each player $i$ is associated with a strategy vector $s_i = [s_i^/, s_i^{\Delta}]$, which defines the strategy $s_i^/$ adopted by the player in pairwise games and  $s_i^{\Delta}$ in $3$-player games.
We focus on social dilemma games, therefore each player can either choose to cooperate (strategy $s_i^{[\cdot]} = C$) or defect ($s_i^{[\cdot]}=D$) in each type of interaction independently, where $[\cdot] \in \{/, \Delta\}$.

\subsection{Payoffs}\label{sec:payoff}

The payoffs defining the symmetric 2-player game can be conveniently represented as a payoff matrix $M^/$:
\begin{align}
M^/ = \kbordermatrix{
    & C & D \\
    C & R^/,R^/ & S^/,T^/ \\
    D & T^/,S^/ & P^/,P^/ 
  } ,
\end{align}
where the first entry in each cell denotes the payoff earned by the row player, while the second entry is the payoff for the column player. $R^/$ denotes the reward for mutual cooperation, while $P^/$ denotes the penalty for mutual defection. A cooperator earns a payoff $S^/$ (namely \emph{sucker}'s payoff) against a defector, while the defector earns a payoff $T^/$ (\emph{temptation} to defect) when facing a cooperator. The relative order of these payoffs determines the type of 2-player social dilemmas. 
In particular for our analysis, we will study prisoner's dilemma games, defined by the ordering $T^/ > R^/ > P^/ > S^/$. To reduce the number of parameters in the game, we fix $R^/ = 1$ and $P^/ = 0$ to define the limit of the parameters as is typically done in the literature \cite{szabo_evolutionary_2007, guo_evolutionary_2021, wang_impact_2023, civilini_explosive_2024, wang2024mixing}.

Analogously to the case of pairwise games, the payoffs in a $3$-player game are determined by the combination of strategies of the players. However, with three players, the payoffs depend on the three strategies of the players considered simultaneously. Therefore, in the case of $3$-player games the higher-order game payoff structure can be represented as a $2 \times 2 \times 2$ payoff cube $M^{\Delta}$, such that each axis denotes the possible strategies of each player \cite{civilini_explosive_2024}. For the sake of readability, we represent this payoff cube as two $2 \times 2$ payoff matrices $M^{\Delta}(C)$ and $M^{\Delta}(D)$ stacked on top of each other (fig.~\eqref{fig:illustration}):

\begin{align}
\begin{split}
M^{\Delta}(C) = \kbordermatrix{
    & C & D \\
    C & R^{\Delta},R^{\Delta},R^{\Delta} & G^{\Delta},T^{\Delta},G^{\Delta} \\
    D & T^{\Delta},G^{\Delta},G^{\Delta} & W^{\Delta},W^{\Delta},S^{\Delta}
  }, \\
M^{\Delta}(D) = \kbordermatrix{
    & C & D \\
    C & G^{\Delta},G^{\Delta},T^{\Delta} & S^{\Delta},W^{\Delta},W^{\Delta} \\
    D & W^{\Delta},S^{\Delta},W^{\Delta} & P^{\Delta},P^{\Delta},P^{\Delta} 
  },
\end{split}
\end{align}

where $M^{\Delta}(C)$ and $M^{\Delta}(D)$ respectively denote the payoff matrices when the third player is cooperating or defecting. 
Each entry of the matrix is a $3$-tuple with the payoffs of player 1 (row), player 2 (column), and player 3 respectively. As for pairwise games, $R^{\Delta}$ and $P^{\Delta}$ denote respectively the payoffs for mutual cooperation and mutual defection among 3 players. $T^{\Delta}$ is the higher-order temptation payoff for deviation from mutual cooperation by defecting and $S^{\Delta}$ is the higher-order sucker's payoff for deviation from mutual defection by cooperating. However, the 3-player game introduces some new payoffs which do not have any counterpart in the 2-player game. $G^{\Delta}$ denotes the payoff of a cooperator in a group with two defectors, i.e. for the strategy profile $\{C, C, D\}$ and all its permutations. On the other hand $W^{\Delta}$ denotes the payoff of a defector playing against two cooperators i.e. corresponding to $\{D, D, C\}$ and all its permutations \cite{civilini_explosive_2024}.

\vspace{1em}
\paragraph*{Higher-order social dilemma:} A social dilemma is a type of collective action problem where there is a tension between personal benefit and collective good.
For instance, a $2$-player Prisoner's dilemma represents a social dilemma since, given the payoff ordering $T^/>R^/>P^/>S^/$ defining the game, the rational choice (or Nash equilibrium) for a player is to choose defection, even though it would be more beneficial (i.e., it would bring a higher payoff) to cooperate for both players. 
However, generalizing the concept of social dilemma to multi-player games is not trivial. Over the years, different definitions \cite{kerr2004altruism, pena_ordering_2016} of social dilemmas in multi-player game representation have been proposed. Here, we adopt the definition proposed by Ref. \cite{ broom_generalized_2019}, according to which the payoffs of a $3$-player game have to satisfy the following conditions to qualify as a social dilemma:

\begin{enumerate}[label=\alph*.]
    \item A focal player benefits when other members cooperate, regardless of its own strategic choice. This leads to the following payoff relationships in our framework:
    \begin{align}
        R^{\Delta} &\geq G^{\Delta} \;\geq S^{\Delta}, \\
        T^{\Delta} &\geq W^{\Delta} \geq P^{\Delta}
    \end{align}
    \item Cooperating mutually yields a greater reward than mutual defection:
    \begin{align}
        R^{\Delta} > P^{\Delta}
    \end{align}
    \item Within a group, defectors receive higher payoffs than cooperators:
    \begin{align}
        T^{\Delta} &> G^{\Delta}, \\
        W^{\Delta} &> S^{\Delta}
    \end{align}
    \item Switching from cooperation to defection results in a higher payoff. In a three-player setting, this translates to:
    \begin{align}
        T^{\Delta} &> R^{\Delta}, \\
        W^{\Delta} &> G^{\Delta}, \\
        P^{\Delta} &> S^{\Delta}
    \end{align}
\end{enumerate}

We characterize the higher-order Prisoner's Dilemma by maintaining the same relative ranking of payoffs as in the standard two-player case:

\begin{align}
    T^{\Delta}>R^{\Delta}>P^{\Delta}>S^{\Delta}
\end{align}

However, in the higher-order game, two additional payoffs, $G^{\Delta}$ and $W^{\Delta}$, are present. If $G^{\Delta} < W^{\Delta}$, all the above conditions (a-d) hold, classifying the game as a \textit{strong} social dilemma. Conversely, when $G^{\Delta} > W^{\Delta}$, condition d is not met, leading to what is known as a \textit{relaxed} social dilemma. In general, the quantity $\alpha=W^{\Delta}-G^{\Delta}$ can be interpreted as a measure of the strength of the higher-order Prisoner's Dilemma (PD). Notably, the strong social dilemma has only full defection ($\{D, D, D\}$) as a Nash equilibrium, whereas in the relaxed social dilemma, besides full defection, the strategy profile $\{C, C, D\}$ also forms a Nash equilibrium \cite{gokhale_evolutionary_2010, hilbe_cooperation_2014, civilini_explosive_2024}.

\vspace{1em}
\paragraph*{Comparability of pairwise and higher-order games:} 

It is crucial that payoffs for hyperedges of different sizes remain comparable so that larger groups do not give an undue advantage (or disadvantage) simply because of their group interaction \cite{capraro_group_2015, pena_group_2018, pereda_group_2019, wang_evolutionary_2024}. For instance, consider a situation where a player cooperates with two cooperating players in two dyadic interactions earning a total payoff of $2R^/$. If instead the player cooperated with the same two cooperating players but in a 3-player interaction, it will earn a payoff of $R^{\Delta}$. Since the evolutionary dynamics typically consists of aggregating payoffs to determine the fitness of strategies, if $2 R^/ \neq R^{\Delta}$, the total payoff earned by a player by interacting with the same players but with different types of interaction (pairwise or higher-order) will be different. Even though larger groups can have synergistic effects \cite{ capraro_group_2015, pena_group_2018, pereda_group_2019, lee_group-size_2023} making cooperation naturally more or less advantageous, we want to focus on emergent collective behaviour in absence of these phenomena. Inspired by this line of thinking, one can imagine that from a player's perspective, a single 3-player interaction is equivalent to two pairwise interactions.
Here, we introduce a new payoff constraint for the pairwise and higher-order payoffs, $R^{\Delta} \sim R^/ + R^/ = 2 R^/$ and similarly for $S^{\Delta} = 2 S^/$, $T^{\Delta} = 2 T^/$, and $P^{\Delta} = 2 P^/$. We tune the values of $G^{\Delta}$ and $W^{\Delta}$ to explore the landscape of higher-order PD based on its strength mentioned above.

Put together, we assign the following payoff entries,
\begin{itemize}
    \item $T^/, R^/, P^/, S^/ = 1.1, 1, 0, -0.1$
    \item $T^{\Delta}, R^{\Delta}, P^{\Delta}, S^{\Delta} = 2.2, 2, 0, -0.2$
    \item $W^{\Delta}=0.7$ and denoting the social dilemma strength as $\alpha = W^{\Delta}-G^{\Delta} \in [-1.4, 0.3]$
\end{itemize}

\subsection{Topological overlap in higher-order networks}

We consider a random-regular hypergraph such that a player $i$ has $k^/$ 2-hyperedges (or pairwise neighbours) and further participates in $k^{\Delta}$ 3-hyperedges (or triangles). For our simulations, we fix $k^/ = 4$ and $k^{\Delta} = 2$ so that each player interacts with maximum 8 (=$k^/+2k^{\Delta})$ other unique players.
We can represent any form of higher-order interactions as a multilayer hypergraph, where each layer consists of interactions of a specific size/order, as illustrated in fig.~\eqref{fig:illustration}. The advantage of such a representation is that we can adapt well-established measures from the theory of multiplex networks to characterize the structure of our system. We are interested in exploring how topological similarity between the pairwise and higher-order interactions impacts the emergence of cooperation in structured populations \cite{battiston_structural_2014, lee_how_2021, anwar_intralayer_2022, zhang_higher-order_2023, presigny_node-layer_2024, malizia2025hyperedge, lamata2025hyperedge}. 

We denote $\mathcal{E}^{\Delta}_{\text{proj}}$ as the set of projection of 3-hyperedges unto their corresponding $2$-hyperedges. In other words, if $e_{lmn} \in \mathcal{E}^{\Delta}$, then $e_{lm}, e_{mn}, e_{ln} \in \mathcal{E}^{\Delta}_{\text{proj}}$. We then define the topological overlap $\omega$ between $\mathcal{E}^/$ and $\mathcal{E}^{\Delta}$ as the size of the intersection between the sets of edges $\mathcal{E}^/$ and $\mathcal{E}^{\Delta}_{\text{proj}}$ normalized by the size of set of edges in the pairwise network \cite{battiston_structural_2014, battiston_determinants_2017, krishnagopal_topology_2023, lamata2024integrating},

\begin{align}
    \omega = \frac{| \mathcal{E}^/ \cap \mathcal{E}^{\Delta}_{\text{proj}} |}{|\mathcal{E}^/|}
\end{align}

Thus, if all the pairs of players playing a higher-order game in $\mathcal{E}^{\Delta}$, also participate in 2-player games in $\mathcal{E}^/$, then $\omega=1$. On the other hand, if none of the pair of players playing a higher-order game is participating together in a 2-player game $\omega = 0$. To fine-tune the topological overlap $\omega$, we start by constructing a random-regular hypergraph with full overlap between pairwise and 3-player interactions/layers (i.e., with $\omega =1$, corresponding to a 3-regular simplicial complex). We then decrease the level of overlap $\omega$ by performing a \emph{criss-cross} rewiring of the pairwise interactions. To do that, we consider the pairwise layer in this hypergraph and find two edges $A-B$ and $C-D$, such that nodes $A, B, C, D$ are all distinct and their corresponding subgraph is disjoint, i.e. there are no links from $A$ or $B$ to $C$ or $D$. We then do a double swap rewiring of the old edges $A-B$ and $C-D$, such that $A-D$ and $B-C$ are now the new edges. The advantage of this rewiring method is that it preserves the degrees of all nodes, and it changes the overlap between the pairwise and higher-order layers of the hypergraph. We also ascertain that the graph remains connected following this edge swap even considering all links only and all 3-hyperedges only. We repeat this procedure multiple times to get the desired level of overlap in the system.

\subsection{Evolutionary dynamics}

We evolve the system by using the Monte Carlo method for stochastic dynamics. At each time step, we select a random player $f$ as the focal player and one of its neighbours $m$ in either of the layers as the model player. They both play with all their neighbours in both orders of interaction and collect payoffs $\pi_f$ and $\pi_m$ respectively. Here, $\pi_f = \pi_f^/ + \pi_f^{\Delta}$ and $\pi_m = \pi_m^/ + \pi_m^{\Delta}$ denote the total payoff from both the orders \cite{battiston_determinants_2017} (see fig \eqref{fig:illustration}). We define the transition probability based on the payoffs as the Fermi function: \cite{szabo_evolutionary_1998, traulsen_stochastic_2006, perc_evolutionary_2013},

\begin{align}
    \Pi_t = \frac{1}{1 + \exp[-w(\pi_m-\pi_f)]},
\end{align}

where $w$ quantifies the noise in the copying process. In the limit $w \rightarrow \infty$, $\Pi_t$ approaches 1 if $\pi_m > \pi_f$ and 0 if $\pi_f > \pi_m$. In the other limit $w \rightarrow 0$, $\Pi_t$ approaches 0.5, independent of the magnitudes of $\pi_f$ and $\pi_m$, thus becoming a random process. We choose an intermediate value of noise, $w = 1/(k^/ + k^{\Delta}) \approx 0.16$ \cite{nowak_emergence_2004, wu_universality_2010, allen_nonlinear_2024}.

To account for the dynamical coupling between interaction orders, we assume that the focal player can imitate the strategies of the model player associated with the two layers. To illustrate with an example, consider the update of the pairwise strategy of a focal player $f$ denoted by $s_f^/$ (fig. \eqref{fig:illustration}). We now assume that the player can copy the pairwise strategy of the model player $m$ denoted by $s_m^/$. However, here we propose that the dynamical coupling between the two layers also enables the focal player to copy the higher-order strategy of the model player denoted by $s_m^{\Delta}$. We denote the probability of imitating the strategy of the model player from the other layer as $p_{\text{switch}}$, controlling the dynamical coupling between the two layers and allowing for inter-order imitation.
The profile of all transition probabilities at a given time step can be summarised as follows,

\begin{eqnarray}
 s_f^/ \rightarrow 
    \begin{cases}
    s_m^/& \text{with } 0.5\cdot (1-p_{\text{switch}})  \cdot \Pi_t\\
    s_m^{\Delta}& \text{with } 0.5\cdot p_{\text{switch}}  \cdot \Pi_t\\ 
    \end{cases}\\
s_f^{\Delta} \rightarrow 
    \begin{cases}
    s_m^{\Delta}& \text{with } 0.5\cdot (1-p_{\text{switch}})  \cdot \Pi_t\\
    s_m^/& \text{with } 0.5\cdot p_{\text{switch}}  \cdot \Pi_t\\
    \end{cases}
\end{eqnarray}

We obtain one full Monte-Carlo Step (MCS) by repeating the above procedure $2\cdot N$ times so that each player gets the opportunity to update both the pairwise and higher-order strategies once on average. The structural and dynamical components of our model are fully visualized in fig.~\eqref{fig:illustration}.

We characterize the pro-social behaviour in our system using several descriptors. We introduce $\rho_0^/$ and $\rho_0^{\Delta}$ as the initial density of cooperators in the pairwise and higher-order layers respectively. Correspondingly, $\rho_0 = \frac{1}{2} [\rho_0^/+\rho_0^{\Delta}]$ denotes the initial mass of cooperators in the system. We always consider $\rho_0^/=\rho_0^{\Delta}=\rho_0 = 0.5$ unless stated otherwise. To quantify the stationary state properties of the system, \textit{i.e.} a state where the observables of the system become time independent, we compute $\rho^/$ as the fraction of nodes cooperating in the pairwise interactions and $\rho^{\Delta}$ as the fraction of nodes cooperating in the higher-order interactions. We denote the overall cooperation level in the system as \cite{wang_evolutionary_2015, battiston_determinants_2017},

\begin{align}
    \rho = \frac{1}{2} \big[\rho^/ + \rho^{\Delta}\big]
\end{align}

Furthermore, $\langle \rho \rangle$ denotes the ensemble average of $\rho$ over $M$ different independent runs.

\section{Results}\label{sec:results}

\begin{figure*}[!ht]
    \centering
    \includegraphics[width=1.85\columnwidth]{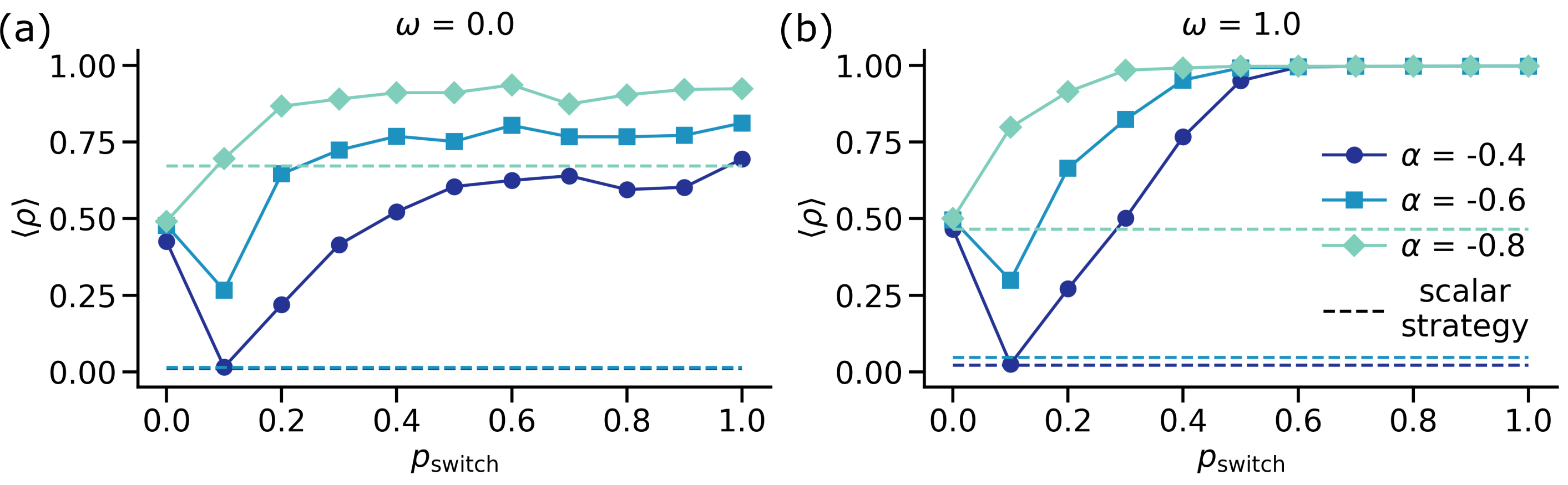}
    \caption{\textbf{Inter-order dynamical coupling mediates cooperation}. Total fraction of cooperative individuals $\langle \rho \rangle$ at the stationary state as a function of $p_{\text{switch}}$ for various values of social dilemma strength $\alpha$ and two values of structural overlap: \textbf{(a)} $\omega=0$ and \textbf{(b)} $\omega=1$. In both panels, the dotted lines denote the reference levels of cooperation if the system consisted of a scalar-strategy profile instead of a 2-dimensional strategy vector for the individuals. Results are shown for the random-regular hypergraph with $N=1500, k^/=4, k^{\Delta}=2$. The initial levels of cooperation in the system are $\rho_0^/=\rho_0^{\Delta}=0.5$. All the results are plotted after averaging over $M=400$ independent runs.}
    \label{fig:p_switch}
\end{figure*}

We analyze the outcome of the evolutionary dynamics of our model by observing the system in the stationary state. We consider a population of $N=1500$ individuals interconnected through edges and hyperedges. Each individual is connected to $k^/=4$ individuals through pairwise interactions. Each individual is also connected to 4 neighbours through $k^{\Delta} = 2$ hyperedges of size 3. We evolve the system using the quasi-stationary method, often used in stochastic processes with absorbing states to find the stable point of the underlying dynamical process \cite{de_oliveira_how_2005, zhou_evolutionary_2010,  faure_quasi-stationary_2014, sander_sampling_2016}. The quasi-stationary method dictates that if the system reaches an absorbing state, which in our case would be full cooperation or defection for either interaction order, the system is reverted (or teleported) to one of the previously visited states with a probability proportional to the time spent by the system in that particular state. We evolve the system for $10^5$ MCS, starting from an equal number of cooperators and defectors for each type of interaction, and look at the average levels of cooperation in the last $10^4$ MCS, defined as the stationary state across $M=400$ independent runs.

\subsection{Inter-order dynamical coupling mediates cooperation}

We first investigate the effect of the dynamical coupling between the different orders of interactions in our model by tuning $p_{\text{switch}}$. Figure~\eqref{fig:p_switch} depicts the stationary state cooperation levels $\langle \rho \rangle$ as a function of $p_{\text{switch}}$ for various values of social dilemma strength $\alpha$ and structural overlap $\omega$ in multi-dimensional strategy systems. To highlight the advantages of our model, we also plot in coloured dotted lines, the cooperation levels in a similar system except with a single/scalar strategy assigned to each node instead of a vector of strategies. In other words, we showcase the difference between the presence and absence of group-size-dependent strategies.  Note that in the case where each node has only one associated strategy, the notion of $p_{\text{switch}}$ is meaningless. First, we observe that depending on the value of $\alpha$, the uni-dimensional scalar-strategy system shows a transition from full defection to high levels of cooperation \cite{civilini_explosive_2024}. Second, we notice that for both values of overlap $\omega=0.0$ and $\omega=1.0$, for $p_{\text{switch}} = 0$, the system displays 50\% cooperative agents. We can understand this behaviour by noticing that the optimum strategy for the pairwise PD game based on our chosen payoff values is pure defection, while for the higher-order game, since $\alpha < 0$, one of the Nash equilibria is to cooperate (see Sec.~\eqref{sec:payoff}) \cite{gokhale_evolutionary_2010, civilini_explosive_2024}.

The situation changes drastically when we increase the dynamical coupling in the system by changing $p_{\text{switch}}$. For very small values of $p_{\text{switch}} < 0.1$, the stationary state is non-trivially dependent on $\alpha$ as well as $\omega$. We see a decrease in cooperation levels for all values of parameters except when the structural overlap is high. However, increasing the dynamical coupling even further ($0 < p_{\text{switch}} < 0.3$), the system shows signs of pro-social behaviour as seen from an increased level of cooperation. Note that this increase is enhanced in the full structural overlap case ($\omega=1.0$). It is crucial to observe that for all sets of parameters, the cooperation levels in the multi-dimensional strategy system are at least comparable to or greater than the same levels in scalar-strategy systems. Increasing the dynamical coupling even more ($p_{\text{switch}}>0.3$), we notice that the cooperation levels saturate. The saturation value is dependent on $\alpha$ for $\omega=0$, but for full structural overlap $\omega=1$, the system displays full cooperation independent of $\alpha$ for a wide range of $p_{\text{switch}}$.

\begin{figure*}[!ht]
    \centering
    \includegraphics[width=2\columnwidth]{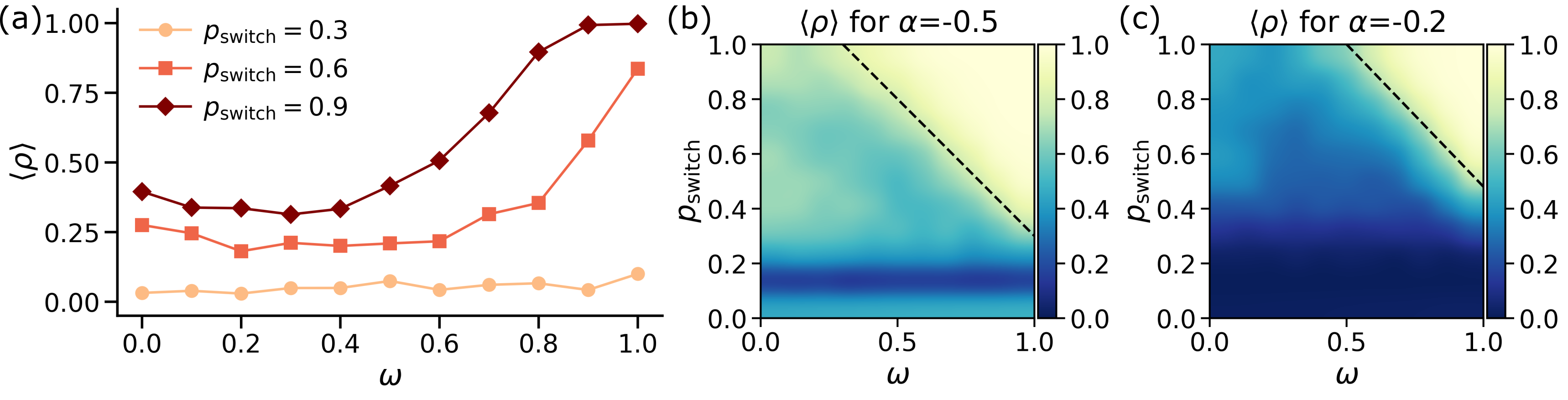}
    \caption{\textbf{Inter-order structural overlap promotes pro-sociality.} \textbf{(a)} Density of cooperators $\langle \rho \rangle$ as a function of structural overlap $\omega$ for various values of $p_{\text{switch}}$ and $\alpha=-0.1$. Heatmap of the stationary state cooperation levels as a function of $\omega$ and $p_{\text{switch}}$ for \textbf{(b)} $\alpha=-0.5$ and \textbf{(c)} $\alpha=-0.2$. Darker shades of blue denote lower levels of cooperation while lighter shades of yellow denote higher values of cooperation. The dotted line separates the area in which very high levels of cooperation ($\langle \rho \rangle > 0.8$) are observed.}
    \label{fig:overlap}
\end{figure*}

To summarize, hypergraph structure does not readily promote pro-social behaviour in the absence of dynamical coupling ($p_{\text{switch}} = 0$). However, when we increase the dynamical coupling to a small non-zero value ($p_{\text{switch}} > 0.2$), the system prompts an increase in cooperative behaviour. This increase is mediated by a nuanced balance between the dynamical coupling strength $p_{\text{switch}}$ and structural overlap $\omega$, where for higher values of overlap, cooperation is heavily preferred. Furthermore, we get higher levels of cooperation in our model compared to the scenario if the agents were described by a scalar strategy.  All in all, our results display a rich interplay between the structural and dynamical components of the system to elevate levels of cooperation.

\subsection{Inter-order structural overlap promotes pro-sociality}

Social behaviour in groups is usually different from behaviour in pairwise interactions \cite{krause_social_2007, capraro_group_2015, pereda_group_2019}. One of the leading hypothesis for this difference is that a group is not merely a sum of its parts, but rather that emergent synergistic effects are manifested in the form of peer pressure which fundamentally change the interactions \cite{christakis_spread_2007, fowler_dynamic_2008}. In this context, it is interesting to understand scenarios where two individuals interact in social contexts with differing number of co-participants. Figure \eqref{fig:overlap} shows how structural overlap between interactions of different orders controls the levels of cooperation in a system. This connection between group interactions and overlap arises because structural overlap determines how influence and behavioural reinforcement propagate across individuals. In systems with high overlap, the same individuals can participate in multiple groups, amplifying peer effects and stabilizing cooperative norms, whereas in low overlap limits such reinforcement, can reduce the persistence of cooperation.

Figure \eqref{fig:overlap} (a) shows the behaviour of the system as a function of tunable topological overlap $\omega$ for various values of dynamical coupling $p_{\text{switch}}$ and a fixed value of social dilemma strength $\alpha=-0.1$. We observe that for small values of $p_{\text{switch}} = 0.3$, there is no effect of the structural overlap on the stationary state cooperation levels. The cooperation levels are low and independent of $\omega$. However, when we increase the dynamical coupling to $p_{\text{switch}}=0.6$, the structural overlap between the different layers starts playing a crucial role. In particular, for high values of $\omega >0.7$, the cooperative strategies are preferentially chosen more and we see elevated levels of cooperation in the system than before. When we tune the dynamical coupling to even higher values ($p_{\text{switch}}=0.9$), we see a systematic increase in cooperation for almost all values of $\omega$. Furthermore, the system shows full cooperation for very high values of $\omega > 0.9$. 

To gain a deeper insight into the interplay between the dynamical coupling ($p_{\text{switch}}$) and structural coupling ($\omega$), we plot a heatmap of stationary state cooperation for $\alpha=-0.5$ (fig.\eqref{fig:overlap} panel b) and $\alpha=-0.2$ (fig.\eqref{fig:overlap} panel c) as a function of both $p_{\text{switch}}$ and $\omega$. We use a black dotted line to separate areas of high cooperation levels ($\langle \rho \rangle > 0.8$) in the phase space. For a more \textit{relaxed} social dilemma ($\alpha = -0.5$ in panel b) we notice that for small values of dynamical coupling, \textit{i.e.} $p_{\text{switch}}<0.3$, topological overlap plays no part in the stationary state of the system. The cooperation levels fluctuate around the $0.5$ for no dynamical coupling, while they drop down to $0.1$ for small values of $p_{\text{switch}}$. However, when we increase the dynamical coupling ($p_{\text{switch}}>0.5$), a high topological overlap promotes cooperation to a greater degree. The role of structural overlap becomes more and more important as we increase the dynamical coupling. In particular, for the highest level of dynamical coupling $p_{\text{switch}}>0.8$, we get full cooperation in the system even for relatively lower values of $\omega \sim 0.5$. When we increase the social dilemma strength to $\alpha = -0.2$ (fig.\eqref{fig:overlap} panel c), we get an overall decrease in cooperation levels. This is expected since increasing the social dilemma strength makes the temptation to defect even stronger. We notice that the region of very high cooperation levels separated by the black dotted line is now much smaller. Additionally, for lower values of dynamical coupling ($p_{\text{switch}}<0.3$), the system displays high levels of defection independent of the level of structural overlap. Note that the white-yellow region persists even beyond $\alpha>0$ (not plotted here), signifying that even for \textit{strong} social dilemmas, high structural overlap and high dynamical coupling can promote cooperation.

\begin{figure*}[!ht]
    \centering
    \includegraphics[width=2\columnwidth]{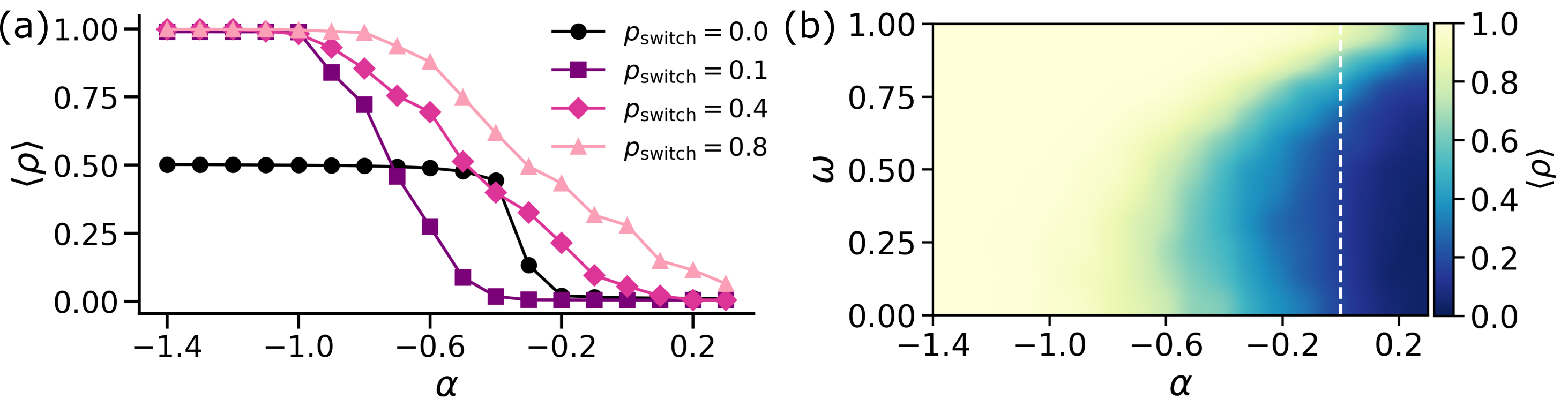}
    \caption{\textbf{Higher-order social dilemma strength modulates cooperative behaviour}. \textbf{(a)} Fraction of agents cooperating in the stationary state as a function of the higher-order social dilemma strength $\alpha$ for various values of dynamical coupling $p_{\text{switch}}$ and $\omega=0.5$. \textbf{(b)} Heatmap showing the pattern of overall cooperation $\langle \rho \rangle$ in the system from low (blue) to high (yellow/white) as a function of topological overlap $\omega$ and strength of social dilemma $\alpha$.}
    \label{fig:dilemma_strength}
\end{figure*}

All in all, our analysis reveals that edge-hyperedge overlap in hypergraphs is one of the major drivers of cooperation in social dilemma situations. The pro-social effects are most visible when the dynamical coupling is also high. These results hint at the importance of reinforcement and peer pressure present in real life to elevate cooperation levels.

\subsection{Higher-order social dilemma strength modulates cooperative behaviour}

We now turn our attention to exploring how changes in the strength of the social dilemma impact the cooperation levels observed in the population. In particular, we are interested in regimes close to $\alpha = 0$, where the higher-order Prisoner's dilemma transitions from being a \textit{relaxed} social dilemma to a \textit{strong} social dilemma.

Figure \eqref{fig:dilemma_strength} (a) showcases the behaviour of the system as a function of $\alpha$ for various values of $p_{\text{switch}}$ and structural overlap $\omega=0.5$. First, we notice that in the absence of dynamical coupling $p_{\text{switch}}=0$ (black line, circular symbols), the system shows half cooperation up to $\alpha=-0.4$ and then transitions to full defection for the rest of the values of $\alpha$. The lack of any dynamical coupling pushes the system to a state characterized by full defection in pairwise games and full cooperation in higher-order games leading to 50\% overall cooperation. However, for high social dilemma strength, we always get full defection which is not desirable. The trend is qualitatively different when we have non-zero values of dynamical coupling present in the system. Most importantly, for very low values of $\alpha < -1$, the system shows full cooperation for all values of $p_{\text{switch}}>0$. However, as we start increasing the social dilemma strength, we notice a fall in the cooperation levels to $\langle \rho \rangle = 0$ for \textit{strong} social dilemmas. This holds true for all values of $p_{\text{switch}}$. The important question to ask in this context is how the cooperation levels drop as we increase the social dilemma strength.

Figure \eqref{fig:dilemma_strength} (a) shows the trend of $\langle \rho \rangle$ for various values of $p_{\text{switch}}$. For $p_{\text{switch}}=0.1$ (purple curve, square symbols), the cooperation drops relatively earlier and the cooperators lose majority ($\langle \rho \rangle < 0.5$) for $\alpha \sim -0.7$. The cooperation levels continue to plummet and we get full defection for values of $\alpha$ that still satisfy the conditions for a \textit{relaxed} social dilemma. When we increase the dynamical coupling to $p_{\text{switch}} = 0.4$ (pink curve, diamond symbols), we see a relatively slower decay of cooperation levels. Furthermore, the cooperation is (almost) always higher than the case with no dynamical coupling. Additionally, we get non-zero values of cooperation even at high social dilemma strengths ($\alpha \sim -0.1$). Finally, when we tune the dynamical coupling to $p_{\text{switch}} = 0.8$ (light pink curve, triangular symbols), we get the slowest decay in cooperation levels. As a consequence, cooperation is always strictly higher than in the scenario with no dynamical coupling. Moreover, we get non-zero values of cooperation even in the \textit{strong} social dilemma strength regimes ($\alpha \geq 0$).

To gain deeper insight into the mechanisms promoting pro-sociality, we plot the stationary state cooperation levels as a function of $\alpha$ and $\omega$ for $p_{\text{switch}}=0.7$ in fig.~\eqref{fig:dilemma_strength} (b). We observe that for low values of social dilemma strengths ($\alpha < -0.9$), we get full cooperation independent of the value of structural overlap $\omega$. This is expected since the temptation to defect is lower for relaxed social dilemmas. However, when we move closer to the limit of \textit{relaxed} social dilemmas denoted by the white dotted line, i.e. $\alpha = 0$, we get some non-trivial patterns in cooperation levels. In particular for $-0.6 < \alpha < 0$, the cooperation levels are intermediate. However, higher structural overlap generally promotes cooperation more for a given value of $\alpha$. For \textit{strong} social dilemmas ($0 < \alpha < 0.3$), the system is mostly dominated by defectors with only 10\% to 20\% population cooperating in the long-time limit. However, even in this case, very high levels of structural overlap seem to sustain cooperation. In particular, for higher structural overlap, the system sustains cooperation for a long range of social dilemma strengths.

Put together, our results display rich emergent behaviour as we tune the strength of the social dilemma. High social dilemma strength tends to inhibit cooperation in the system. However, high levels of structural overlap and large dynamical coupling can sustain pro-social behaviour even in scenarios where cooperation is unfavourable due to a high temptation to defect.

\section{Conclusions}\label{sec:conclusions}

Over the last few years, higher-order interactions have emerged as an important tool to effectively model networked populations \cite{grilli_higher-order_2017, battiston_networks_2020, battiston_physics_2021}. In particular, research on dynamical processes such as epidemic spreading \cite{iacopini_simplicial_2019}, synchronization \cite{gambuzza_stability_2021}, and more recently evolutionary games \cite{alvarez-rodriguez_evolutionary_2021, civilini_evolutionary_2021, guo_evolutionary_2021, civilini_explosive_2024} has showcased multiple instances where the presence of higher-order interactions fundamentally changes the stationary state properties of the system. While a much richer and more nuanced landscape of possibilities has opened up to examine the critical components of these high-dimensional processes, current works have not investigated in detail how the topology and dynamics of higher-order networks shape their evolutionary behaviour.

Here, we proposed a new model for games on hypergraphs and showed how structural and topological features can lead a population towards cooperative behaviours. The key feature of the model is the representation of higher-order interaction structure as a multilayer hypergraph, where each layer represents the group interactions of a given size. The agents are endowed with different strategies depending on the order of the interaction. Such strategies are dynamically coupled through the introduction of the parameter $p_{\text{switch}}$ which allows the agents to imitate the strategies of their neighbours across different orders. Finally, the hypergraph structure is controlled by tuning the overlap $\omega$ between the edges (size = 2) and hyperedges (size = 3) of the hypergraph.

Our model is characterized by a rich and nuanced interplay between the dynamical coupling $p_{\text{switch}}$ and structural overlap $\omega$ across various strengths of social dilemma strength $\alpha$. First, the multidimensional nature of strategic behaviour allowed to promote cooperation. Second, increasing the dynamical coupling between the different orders of interactions increased the overall cooperation levels in the system. Third, higher structural overlaps further promoted cooperation in this multi-dimensional strategy system. Taken together, our results showed that each of the above three components can elevate cooperation, and the effect is further enhanced when these components are tuned simultaneously.

To summarize, we have proposed a new evolutionary model which sheds new light on the drivers of cooperation in social dilemmas on higher-order networks. For future work, we aim to expand our analysis by investigating additional hypergraph structural features such as hyperdegree heterogeneity and correlations. We hope that our work inspires more research on how non-dyadic interactions shape pro-sociality in humans.

\section*{Author contributions}
O.S. and F.B. conceptualized the idea and developed the methodology. O.S. carried out the formal analysis and created the visualizations. A.C. and V.L. provided methodological insights. All authors wrote, reviewed, and edited the paper.

\section*{Acknowledgements}
The computational results presented have been achieved using the Vienna Scientific Cluster (VSC).

\section*{Data and code availability}
The code for reproducing the results presented in the paper can be found at \href{https://github.com/sadekar-onkar/overlap-ho-game}{https://github.com/sadekar-onkar/overlap-ho-game}. The archived code and data can be found at \href{https://zenodo.org/records/15212417}{Zenodo}.

\section*{Funding}
F.B. acknowledges support from the Air Force Office of Scientific Research under award number FA8655-22-1-7025.

\bibliographystyle{apsrev4-1} 
\bibliography{references}

\end{document}